\documentclass[aps,prb,twocolumn,superscriptaddress,longbibliography]{revtex4-2}
\usepackage{amsmath,amssymb}
\usepackage[pdftex]{hyperref,graphicx}
\hypersetup{colorlinks = true, urlcolor = blue, linkcolor = blue, citecolor = blue}
\usepackage{physics}
\usepackage{xcolor}
\usepackage{bm}
\usepackage{pdfpages}

\newcommand{\comment}[1]{{}}

\makeatletter
\AtBeginDocument{\let\LS@rot\@undefined}
\makeatother

\begin{document}
\title{Disorder induced 2D metal-insulator transition in moiré transition metal dichalcogenide multilayers}
\author{Seongjin Ahn}
\affiliation{Condensed Matter Theory Center and Joint Quantum Institute, Department of Physics, University of Maryland, College Park, Maryland 20742, USA}
\author{Sankar Das Sarma}
\affiliation{Condensed Matter Theory Center and Joint Quantum Institute, Department of Physics, University of Maryland, College Park, Maryland 20742, USA}

\begin{abstract}
We develop a minimal theory for the recently observed metal-insulator transition (MIT) in two-dimensional (2D) moiré multilayer transition metal dichalcogenides (mTMD) using Coulomb disorder in the environment as the underlying mechanism. In particular, carrier scattering by random charged impurities leads to an effective 2D MIT approximately controlled by the Ioffe-Regel criterion, which is qualitatively consistent with the experiments. We find the necessary disorder to be around $5$-$10\times10^{10}$cm$^{-2}$ random charged impurities in order to quantitatively explain much, but not all, of the observed MIT phenomenology as reported by two different experimental groups. Our estimate is consistent with the known disorder content in TMDs.
\end{abstract}

\maketitle
\section{Introduction}
Extensive recent experimental \cite{jinStripePhasesWSe22021,wangCorrelatedElectronicPhases2020,huangCorrelatedInsulatingStates2021, reganMottGeneralizedWigner2020, tangSimulationHubbardModel2020, xuCorrelatedInsulatingStates2020, ghiottoQuantumCriticalityTwisted2021, liContinuousMottTransition2021} and theoretical \cite{wuHubbardModelPhysics2018,  panQuantumPhaseDiagram2020, panBandTopologyHubbard2020, panInteractionDrivenFillingInducedMetalInsulator2021, morales-duranMetalinsulatorTransitionTransition2021, zangHartreeFockStudymoire2021, hsuSpinvalleyLockedInstabilities2021, devakulMagicTwistedTransition2021} works have established the half-filled (i.e. one electron or hole per moiré unit cell) 2D mTMD systems to be strongly correlated insulators with an interacting Mott-Hubbard tight binding description providing a reasonable starting point. In particular, the half-filled system, in both homo- and hetero-bilayer mTMDs, is insulating with a finite charge gap in conflict with the noninteracting band description predicting the half-filled case to be a metal. Theoretical descriptions using Mott-Hubbard model's ground state and the associated charge gap \cite{panInteractionDrivenFillingInducedMetalInsulator2021}.  
 Tunability of the mTMD properties by doping or applying external electric field makes mTMD an attractive semiconductor platform for studying correlation effects \cite{jinStripePhasesWSe22021,wangCorrelatedElectronicPhases2020,huangCorrelatedInsulatingStates2021, reganMottGeneralizedWigner2020, tangSimulationHubbardModel2020, xuCorrelatedInsulatingStates2020, ghiottoQuantumCriticalityTwisted2021, liContinuousMottTransition2021}.
 
In the current work, motivated by extensive recent experimental transport measurements \cite{ghiottoQuantumCriticalityTwisted2021,liContinuousMottTransition2021}, we focus on half-filling and the region of doping around half-filling, and ask what happens if the insulating phase is somehow suppressed leading to a transition to an effective metal.
Recent experiments \cite{ghiottoQuantumCriticalityTwisted2021,liContinuousMottTransition2021} have investigated mTMD transport in the presence of doping (i.e., slightly away from half-filling) and in the presence of an applied electric field at half-filling normal to the 2D mTMD layers which suppresses the Mott gap. This doping or field induced zero-temperature insulator to metal transition at half-filling is the subject of our work.

Although the physics of doping and applied field should be very different from a theoretical perspective, they lead to remarkably similar experimental transport phenomenology with the insulator eventually becoming a metal at finite doping and/or finite electric field, manifesting similar temperature-dependent resistivity behavior in both cases (one case, where the filling changes due to doping at fixed field, and the other case, where the filling is fixed at half, but the charge gap is suppressed to zero by the applied field). In both cases, there seems to be a continuous transition or a crossover from the insulating activated temperature dependent resistivity to a metallic resistivity at some characteristic (sample dependent) doping or field. In fact, the two transport phenomenologies under doping or electric field are qualitatively identical, although the actual values of the crossover resistivity at the transition in the two experiments differ. 
This crossover resistance loosely separating the effective metal from the effective insulator is not a sharp universal quantity, and indeed this characteristic crossover resistance differs quite a bit between the homobilayer \cite{ghiottoQuantumCriticalityTwisted2021} and heterobilayer \cite{liContinuousMottTransition2021} mTMDs although both are parametrically of $O(h/e^2)$.
The goal of the current work is to provide a unified explanation for the observed 2D MIT using the physically appealing picture that the insulator-to-metal crossover happens because of disorder in the TMD environment arising from the unintentional presence of random charged impurities. The dominance of disorder in the experimental mTMD samples \cite{ghiottoQuantumCriticalityTwisted2021,liContinuousMottTransition2021} is obvious from the extracted very low maximum experimental low-temperature mobilities ($<4000$ $\mathrm{cm}^2/\mathrm{Vs}$) in the metallic regime. 
 
We note that without disorder the system should be a metal the moment it is very slightly doped away from half-filling \cite{panInteractionDrivenFillingInducedMetalInsulator2021}, but experimentally the metallic behavior emerges only at a finite filling of $\sim$1.1 electrons (or $\sim$0.9 holes) per moiré unit cell, i.e., at a finite critical carrier density $\sim 5\times10^{11}\mathrm{cm}^{-2}$. Similarly, the half-filled system should be a perfect metal at $T=0$ the moment the Mott gap vanishes, but experimentally the emergent metallic phase at the transition has a very high resistivity of $\sim$10,000-50,000 ohms, and the metallic resistivity continues to decrease monotonically as the applied field increases, which is not explicable using the simple Mott-Hubbard picture since nothing should happen \cite{senthilTheoryContinuousMott2008} once the Mott insulator has become a metal with zero gap. Both of these features, however, make sense if we invoke disorder in the environment arising from the presence of random quenched charged impurities. In the doped case, a small carrier density at a low filling away from half-filling is unable to overcome the impurity effects as the electrons suffer from strong localization (or percolation) due to the background disorder, and only when the carrier density is approximately equal to the charged impurity density, metallic transport is possible because the Coulomb disorder then gets screened out by carrier screening. In the half-filled case under an applied field suppressing the Mott gap, again the system is unable to manifest metallic conduction until the metallic carriers in the lower Hubbard band can screen out the Coulomb disorder effect very similar to the situation in regular semiconductors (e.g., Si) where at low doping of the conduction band, no metallic conduction happens unless a threshold voltage is applied to overcome strong localization (or percolation) effects. Thus, in both situations the nature of the insulator to metal transition in the transport behavior is qualitatively similar as in both cases the carriers (electrons or holes) must first screen out Coulomb disorder overcoming strong localization (or percolation) in order to produce metallic transport behavior. Thus, a finite carrier density is necessary for the MIT to happen leading to a continuous crossover rather than an abrupt transition -- just the mere disappearance of the Mott gap is insufficient to produce metallic conduction because of the presence of quenched disorder in TMDs.
 
\begin{figure}[!htb]
  \centering
  \includegraphics[width=\linewidth]{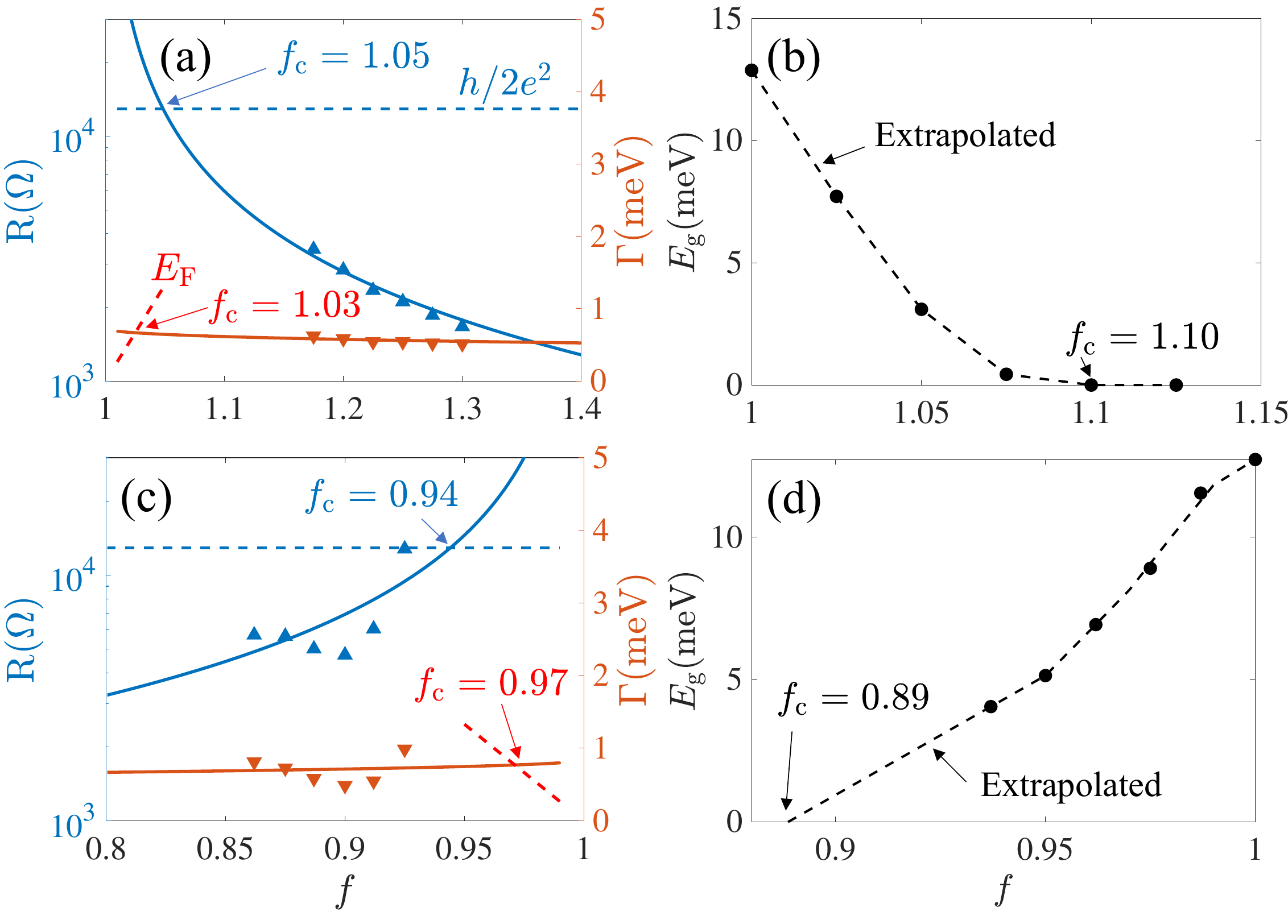}
  \caption{Plot of resistivity $R$, disorder-induced level broadening $\Gamma=\hbar/2\tau$, and the activation gap $E_\mathrm{g}$ (extracted from the experimental activated transport) as a function of the filling factor $f$ for (a),(b) the electron-doped and (c),(d) hole-doped mTMD samples from the Cornell group \cite{Mak}. (a),(c) The low-temperature experimental resistivity (blue, upward-pointing triangle) and the corresponding $\Gamma$ (red, downward-pointing triangle) along with the best fit resistivity curve (solid) obtained using the Boltzmann transport theory with (a) $n_\mathrm{i}=8.71\times 10^{10} \mathrm{cm}^{-2}$ for the electron-doped sample and (c) $n_\mathrm{i}=1.01\times 10^{11} \mathrm{cm}^{-2}$ for the hole-doped sample. The dashed lines represent the resistance quantum $h/2e^2$ (blue), and the Fermi energy $E_\mathrm{F}$ (red). (b),(d) The activation gap extracted from the measured resistivity (dot) on the insulating side. The dashed line is the best linear extrapolate of the activation gap. The values of the critical filling factor $f_c$ are obtained from three independent criteria: $R(f_c)=h/2e^2$ (blue), $\Gamma(f_c)=E_\mathrm{F}(f_c)$ (red), and $E_\mathrm{g}(f_c)=0$ (black). }  
  \label{fig:1}
\end{figure}

\begin{figure}[!htb]
  \centering
  \includegraphics[width=\linewidth]{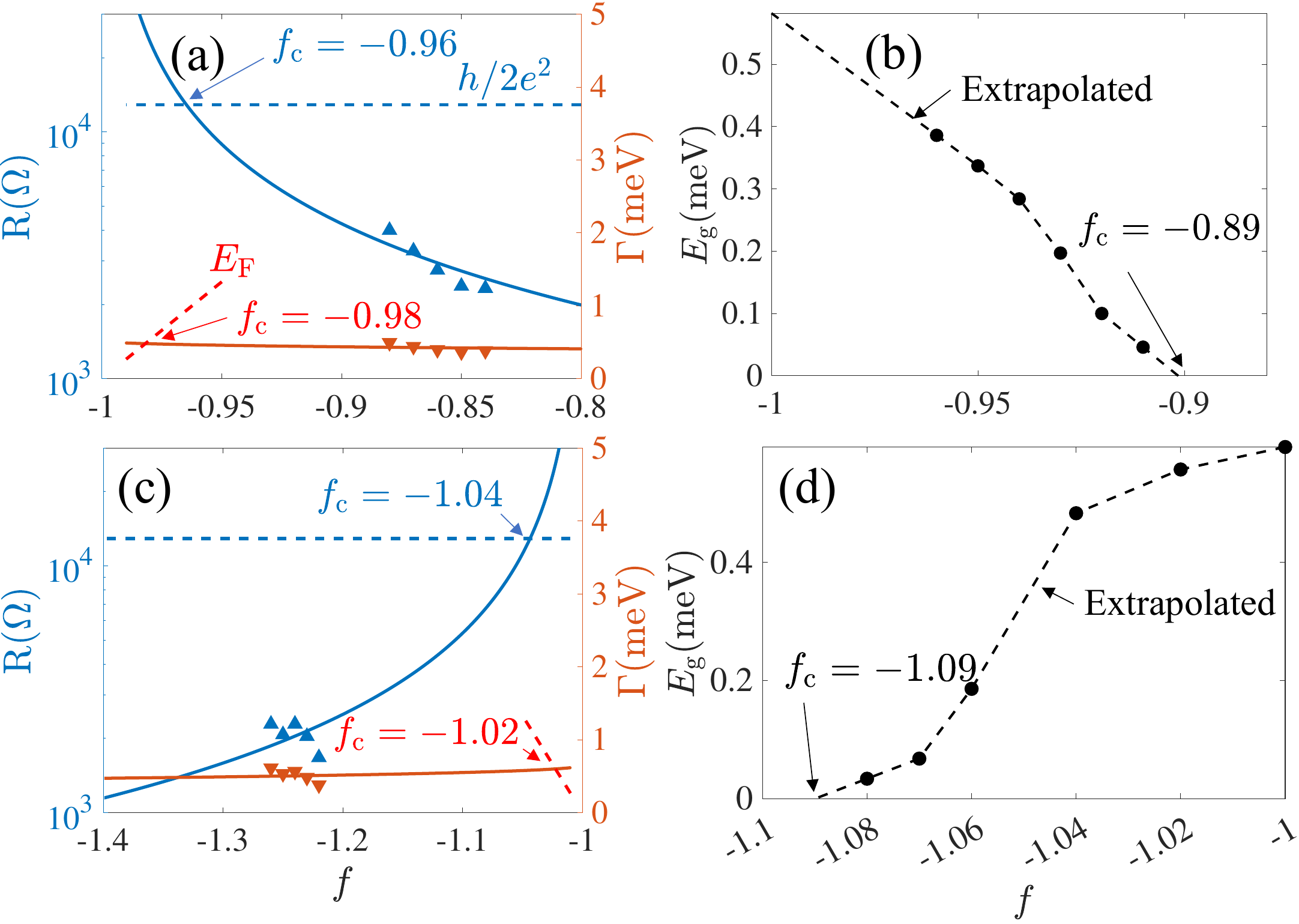}
  \caption{Same as Fig.~\ref{fig:1} but for the (a),(b) electron-doped and (c),(d) hole-doped mTMD samples from the Columbia group \cite{ghiottoQuantumCriticalityTwisted2021}. The experimental low temperature resistivity data are extracted from Fig.~2 of Ref.~\cite{ghiottoQuantumCriticalityTwisted2021}. The best fit is obtained with $n_\mathrm{i}=6.20\times 10^{10} \mathrm{cm}^{-2}$ and $n_\mathrm{i}=7.80\times 10^{10} \mathrm{cm}^{-2}$ for the (a) electron-doped and (c) hole-doped samples, respectively.  }
  \label{fig:2}
\end{figure}

In our theory, we assume a background random charged impurity density $n_\mathrm{i}$ in the system and calculate the metallic resistivity at $T=0$ for a carrier density of $n$, where $n$ (but not $n_\mathrm{i}$) is known experimentally. We employ the Boltzmann transport theory for obtaining the metallic resistivity by calculating the effective screened Coulomb scattering limited transport relaxation time $\tau (n, n_\mathrm{i})$. 
Our comparison to the experiments \cite{ghiottoQuantumCriticalityTwisted2021, liContinuousMottTransition2021, Mak} involves an extrapolation of the measured low-temperature metallic resistivity to $T=0$. For some datasets [Fig.~\ref{fig:1}(c) and Fig.~\ref{fig:2}(c)], the extrapolation is not reliable due to large measurement errors at low temperatures. In such cases, we alternatively present results obtained using finite temperature resistivities less prone to error in the Supplementary Information \cite{supp}, leading to the same conclusion.

We identify the 2D MIT crossover point by using three different criteria, one on the insulating side and two on the metallic side. The crossover point from the insulating side can be determined simply by analyzing the experimental activated resistivity on the insulating side, and extrapolating the activation gap to zero. From the metallic side, we identify the point where the Ioffe-Regel criterion for strong localization (also for percolation \cite{dassarmaTwodimensionalMetalinsulatorTransition2014a}), $\hbar/2\tau(n)=E_\mathrm{F}(n)$, where $E_\mathrm{F}(n)$ is the density-dependent Fermi energy of the metal, is satisfied, defining the lowest carrier density allowed for the metallic transport. We also identify the transition by directly calculating the effective $T=0$ resistivity of the system and equating that to $h/2e^2$, providing a second definition of the MIT crossover from the metallic side. Amazingly, not only do these three definitions of the MIT crossover density agree with each other approximately, our theory also correctly reproduces the full density dependence of the low-temperature metallic resistivity in the experiments! Our Boltzmann theory is limited to the metallic regime and becomes gradually quantitatively worse approaching the MIT crossover point, but the theory does not fail. It has been shown in the vast previous literature on the MIT in semiconductors that the theory remains qualitatively valid approaching the transition from the metallic side \cite{ dassarmaTwodimensionalMetalinsulatorTransition2013, shabaniApparentMetalinsulatorTransition2014, dassarmaScreeningTransport2D2015, manfraTransportPercolationLowDensity2007, tracyObservationPercolationinducedTwodimensional2009, dassarmaTwodimensionalMetalinsulatorTransition2014a}.
Agreement between our theory and the detailed experimental density dependent metallic resistivity is our most important result, establishing the disorder driven crossover to be the correct physics underlying the observed MIT phenomena.
We emphasize that our theory depends only on just one parameter, the impurity density $n_\mathrm{i}$. It is therefore particularly noteworthy that the impurity density necessary for the agreement between our theory and experiment is extremely reasonable, being close to the expected charged impurity density in the TMD environment \cite{Mak}. 
 

\section{Theory and Results}
The zero-temperature resistivity is given by the simple Drude relation $R=m/ne\tau$ where $\tau$ is the scattering time at the Fermi surface.
Within the Boltzmann theory of transport using the relaxation time approximation, the scattering time is given by
\begin{equation}
\begin{aligned}
    \frac{1}{\tau_{\bm k}}&=\frac{2\pi n_\mathrm{i}}{\hbar} \int  \frac{d^2 k'}{(2\pi)^2} \left| \frac{v^{(c)}_q}{\varepsilon_q} \right|^2 \delta(E_{\bm k}-E_{\bm k'})
     (1-\cos{\theta_{\bm k \bm k'}}),
    \label{eq:single_particle_relaxation_time}
\end{aligned}
\end{equation}
where $n_\mathrm{i}$ is the background charged impurity density, $q=\left|\bm k - \bm k' \right|$, $\theta_{\bm k, \bm k'}$ is the scattering angle between $\bm k$ and $\bm k'$, $v^{(c)}_q=2\pi e^2/\kappa q$ is the matrix element of the Coulomb scattering potential between an electron and a charged impurity, $\kappa=5$ is the background lattice dielectric constant for TMD, $\varepsilon_q$ is the 2D screening dielectric function for TMD using the appropriate effective mass which we set $m=0.45$ (in units of free electron mass) \cite{Mak} unless otherwise noted, and $E_{\bm k}=\hbar^2\left|\bm k\right|^2/2m$ is the usual 2D parabolic energy dispersion. Note that we use the effective mass approximation allowing us to ignore the moiré band structure complications which are certainly nonessential since we use the experimentally measured effective mass. Here the 2D screening function $\varepsilon_q$ is given by $\varepsilon_q=1-v^{(c)}_q\Pi^0_q$ \cite{sternPolarizabilityTwoDimensionalElectron1967}, where $\Pi^0_q$ is the noninteracting static polarizability written as
\begin{align}     \label{eq:polar}
    \Pi^0_q=-\frac{m}{\pi \hbar^2}
		\left[1 - \Theta(q-2k_\mathrm{F})\frac{\sqrt{q^2- 4k^2_\mathrm{F} }}{q} \right].
\end{align}


We first consider the doped case, where the resistivity for mTMD is measured as a function of the carrier density around half-filling $f=1$. This is the situation where our theory applies in a straightforward manner since infinitesimal doping away from $f=1$ should produce a metal, but the experimental observation is that the system remains insulating up to a critical finite filling away from $f=1$. For the doped case, we assume that only a partial number of carriers of $n-n_\mathrm{M}$ occupying the upper Hubbard band contribute to the transport, where $n$ is the carrier density and  $n_\mathrm{M}=5\times10^{12} \mathrm{cm}^{-2}$ is the moiré density corresponding to the half-filling.
In Figs.~\ref{fig:1}(a) and (c), we plot the experimental resistivity from the group at Cornell University on the metallic side as a function of the filling factor \cite{Mak} along with the best fit obtained from our transport model. 
We emphasize that our 
fit is in excellent agreement with the measured resistivity despite using only one fitting parameter, the impurity density $n_\mathrm{i}$. We note that the experimental $\Gamma$ varies as a function of the filling factor, implying that short range disorder is not an important resistivity limiting mechanism \cite{dassarmaUniversalDensityScaling2013, Mak}.
The figures also show that the values of the crossover or critical filling factor $f_c$ evaluated according to two independent criteria for predicting the MIT crossover [$R(f_c)=h/2e^2$ and $\Gamma(f_c)=E_\mathrm{F}(f_c)$] differ by only a few percent, which is a very good agreement considering that the Ioffe-Regel criterion is only a crossover criterion valid typically over a factor of 2-3 (Note that the theoretical equivalence between these two criteria is expected, but the important point is that both lead here to agreement with experimental measurements).
On the insulating side [Figs.~\ref{fig:1}(b) and (d)], we assume that the transport is purely activated with no metallic contribution, and extract the energy gap $E_\mathrm{g}$ from this activation behavior manifested in the experimental data. The activation gap gradually decreases as one moves away from half-filling, and eventually vanishes at some critical filling factor $f_c$. Using the linear extrapolation of the extracted activation gap (dashed line) from the experimental data, we estimate the critical filling factor to be $f_c=1.10$ and $f_c=0.89$ (from the insulating side) for the electron- and hole-doped samples, respectively. The values of $f_c$ somewhat depend on the extrapolation scheme used, but we confirm that the difference is not large enough to alter the conclusion. We emphasize that these values of $f_c$ obtained using the experimental data on the insulating side are approximately equal to those obtained from the criteria for the calculated metallic resistivity shown in Figs.~\ref{fig:1}(a) and (c). 
In Fig.~\ref{fig:2}, we apply the same analysis to TMD samples from the Columbia University group (Fig.~2 of Ref.~\cite{ghiottoQuantumCriticalityTwisted2021}). 
Note that the fit using our transport model reproduces well the experimental resistivity with the best-fit impurity density $n_\mathrm{i}$ being in an expected and reasonable range ($\sim10^{10}$-$10^{11} \mathrm{cm}^{-2}$). Similar to the results for the Cornell doped data, we find that the critical filling factors $f_\mathrm{c}$ obtained from the three different criteria, two on the metallic side and one on the insulating side, are in good agreement as well as being in agreement with the Columbia data. Also it is noteworthy that the two best-fit impurity densities for electron-doped and hole-doped mTMD samples are in agreement.

\begin{figure}[!htb]
  \centering
  \includegraphics[width=\linewidth]{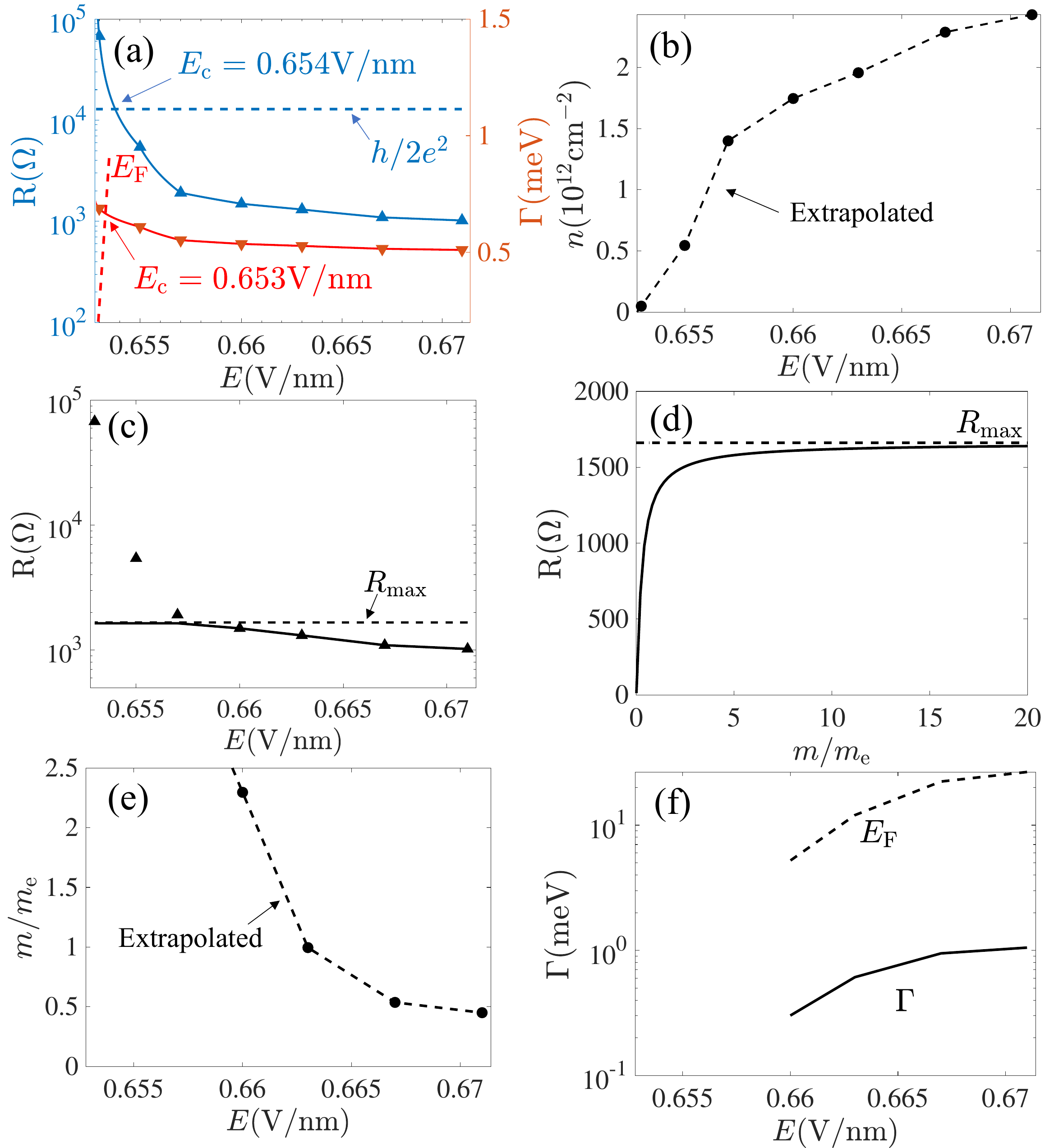}
  \caption{Best fit results to the low-temperature experimental resistivity (triangles, extracted from Fig.~2 of Ref.~\cite{liContinuousMottTransition2021}) of undoped(i.e., $f=1$) mTMD samples under varying electric field from the Cornell group. (a) Best fits obtained assuming that the carrier density $n$ varies with the applied electric field $E$, and (b) the corresponding estimated $n$ from the fitting. The theoretically calculated $R$ and $\Gamma$ plotted as a continuous function of $E$ (solid lines) are computed using the linear interpolation (or extrapolation) of the estimated carrier density $n$ (dot) shown in (b). $E_\mathrm{c}$ is the critical electric field that satisfies the following criteria: $R(E_\mathrm{c})=h/2e^2$ (blue), $\Gamma(E_\mathrm{c})=E_\mathrm{F}(E_\mathrm{c})$ (red). Here we use the impurity density $n_\mathrm{i}=8.71\times10^{10}\mathrm{cm}^{-2}$, which is obtained in the previous analysis for the resistivity of the electron-doped sample from the Cornell group under varying doping density [Fig.~\ref{fig:1}(a)]. 
  (c)-(f) Best fits obtained assuming that the effective mass $m$ varies as a function of $E$ but the carrier density remains fixed at the moiré density (i.e., $n=5\times 10^{12}\mathrm{cm}^{-2}$). Here the impurity density is set to $n_\mathrm{i}=2.05\times 10^{11}\mathrm{cm}$, which is obtained by fitting the most metallic experimental resistivity deep in the metallic regime (i.e., the one for $E=0.671\mathrm{V/nm}$) using the same effective mass as in the previous analysis for doped samples. (c) The solid line represents the best theoretical fit, which fails in the regime $R>R_\mathrm{max}$ because in our Boltzmann model $R$ is bounded from above by $R_\mathrm{max}$ as seen in (d)
  Here $m_\mathrm{e}$ appearing in the horizontal axis is the bare electron mass. (e),(f) Plot of (e) the estimated $m$ from the fitting, which rapidly increases approaching the MIT (i.e., $E\rightarrow E_\mathrm{c}$), (f) the level broadening $\Gamma$ (solid) and the Fermi energy $E_\mathrm{F}$ (dashed) as a function of $E$ in the regime where the fitting is successful (i.e., $R<R_\mathrm{max}$). This figure is for the Device 1 in Ref.~\cite{liContinuousMottTransition2021}-- the same results for Device 2 of Ref.~\cite{liContinuousMottTransition2021} are given in Fig. S1 in the Supplementary Information \cite{supp}.
  }
  \label{fig:3}
\end{figure}

In Fig.~\ref{fig:3}, we consider the mTMD transport experiment from the Cornell group in the presence of the applied out-of-plane electric field $E$ at a fixed filling factor $f=1$ (Fig.~2 of Ref.~\cite{liContinuousMottTransition2021}). 
For the electric-field varying case, we assume all $n_\mathrm{M}$ carriers contribute to the transport unlike the doped case since the applied electric field closes the Mott gap. Motivated by the similarity between the experimentally observed transport phenomenologies under varying doping and electric field, here we first assume that the carrier density $n$ changes with the applied electric field $E$ in the metallic regime. With a fixed impurity density $n_\mathrm{i}=8.71\times 10^{10} \mathrm{cm}^{-2}$ already estimated in the previous analysis for the electron-doped sample from the Cornell group [Fig.~\ref{fig:1}(a)], we evaluate the carrier density dependence on the electric field as shown in Fig.~\ref{fig:3}(b) by fitting separately each of the experimental resistivity measured at different electric fields to our transport model with $n$ being the only tuning parameter. It is worth noting that, as in the doped case, we use only one fitting parameter (the carrier density $n$), but the obtained best-fit carrier densities are within a reasonable range, being smaller than the full moiré density ($\sim 5\times 10^{12}\mathrm{cm}^{-2}$). Using the interpolated (or extrapolated) relation between the carrier density and the electric field (dashed line), we plot in Fig.~\ref{fig:3}(a) 
$R$ and 
$\Gamma$ as a continuous function of the applied electric field, which fit very well the experimental data.  Similar to the doped case, we calculate the critical electric field satisfying the criteria for predicting the MIT crossover on the metallic side, i.e., $R(E_\mathrm{c})=h/2e^2$ and $\Gamma(E_\mathrm{c})=E_\mathrm{F}(E_\mathrm{c})$. Remarkably, we find that the values of $E_\mathrm{c}$ from the two different criteria are in very close agreement within $\pm 0.001 \mathrm{V/nm}$. This implies that the MIT under varying electric field originates from the same physical mechanism (i.e., disorder-driven MIT) as the doped case. 
Note that the $R_\mathrm{c}$ we get in Fig.~\ref{fig:3}(a) is smaller than the experimental $R_\mathrm{c}\sim45$ k-ohms in Ref. \cite{liContinuousMottTransition2021}, which we attribute to percolation effects beyond the Ioffe-Regel strong localization physics \cite{dassarmaTwodimensionalMetalinsulatorTransition2013}.
Note that the carrier density rapidly decreases further below the value of $n_\mathrm{M}$ near the critical electric field in contrast to the experimental observation \cite{liContinuousMottTransition2021} where the carrier density is fixed to $n_\mathrm{M}$, implying that our theory neglecting correlation effects is not sufficient for describing physics near the MIT transition.

A more meaningful assumption \cite{liContinuousMottTransition2021, Mak} for the undoped half-filled case under an electric field is, therefore, that the carrier density beyond the MIT in the metallic regime remains constant at the moiré density of $n\sim 5\times10^{12} \mathrm{cm}^{-2}$, but the carrier effective mass varies with $E$ because of correlation effects and the modification of the moiré hopping induced by the $E$-field. In Fig.~\ref{fig:3} panels (c)-(f), we show our theoretical results for a varying mTMD mass at the fixed moiré carrier density, comparing with the experimental results, getting good agreement between the measured resistivity and the theory using a fixed $n_\mathrm{i}$ and a variable effective mass. There is, however, a serious difference between Figs.~\ref{fig:3} panels (a),(b) (varying carrier density) and panels (c)-(f) (varying effective mass). For varying mass and fixed density, the maximum possible metallic resistivity is bounded from above in our Boltzmann theory by $R_\mathrm{max} =(\pi h/2e{^2}) (n_\mathrm{i}/n) \sim 1700$ ohms [see Fig.~\ref{fig:3}(d)] for our best fit impurity density achieving agreement with the experimental data. This limit exists in our theory by virtue of the strong screening in the large `$m$' limit. This $R_\mathrm{max}$ is substantially lower (by more than a factor of 20) than the nominal $R_\mathrm{c}\sim5\times10{^4}$ ohms observed in the experiment. The existence of $R_\mathrm{max}$ in the theory means that the Ioffe-Regel criterion is not meaningful for the varying mass situation as is obvious in Fig.~\ref{fig:3}(f). Our theory provides excellent agreement with the experimental data for $R<R_\mathrm{max}$, but fails in the regime $R>R_\mathrm{max}$ \cite{liContinuousMottTransition2021}.
Note that this discrepancy cannot be fixed by arbitrarily enhancing $n_\mathrm{i}$ by a factor $>20$ since such a large $n_\mathrm{i}$ would simply be inconsistent with the experimental metallic resisitivity for $R<R_\mathrm{max}$. We believe that our theory fails in the high resistivity metallic regime $R_\mathrm{max}<R<R_\mathrm{c}$ where both insulating and metallic phases coexist near the Mott transition, and the correct physical picture is a percolation picture \cite{dassarmaTwodimensionalMetalinsulatorTransition2013, dassarmaTwodimensionalMetalinsulatorTransition2014a}, rather than the Ioffe-Regel criterion [see Fig.~\ref{fig:3}(f)]. We note that our finding of a divergent effective mass on the metallic side below the Mott transition is consistent with the experiment \cite{liContinuousMottTransition2021} where the carrier effective mass diverges approaching the MIT from the metallic regime. 

\section{Conclusion}
We show that the reported metal-insulator transition in mTMD layers is likely to be a screened Coulomb disorder driven Ioffe-Regel crossover where the free carriers are strongly localized at low densities. Increasing the carrier density screens out the disorder leading to the emergence of a metallic resistivity. Our transport theory provides a quantitatively accurate zeroth order description of the transition at the correct carrier density and the appropriate critical resistance for doped samples, but for undoped samples, our theory fails close to the Mott transition where a percolation picture applies. The qualitative difference between the MIT for the half-filled (doped) case is that there is (not) a gap involved in the crossover, and the Ioffe-Regel criterion strictly applies in the absence of (or far away from) the gap.
An important experimentally verifiable prediction is that increasing (decreasing) random disorder in the sample would lead to an increase (decrease) in the critical filling/field, suppressing (enhancing) the metallic regime. 

\section{Acknowledgement} \label{sec:acknowledgement}
We gratefully thank Professor K. F. Mak for communicating unpublished data on doped samples and for helpful correspondence on the manuscript. This work is supported by the Laboratory for Physical Sciences.



\section*{Supplementary material}

Seongjin Ahn[1] and Sankar Das Sarma[1]

[1]*Condensed Matter Theory Center and Joint Quantum Institute,
Department of Physics, University of Maryland, College Park, Maryland 20742, USA*

## FIT AT FINITE TEMPERATURE

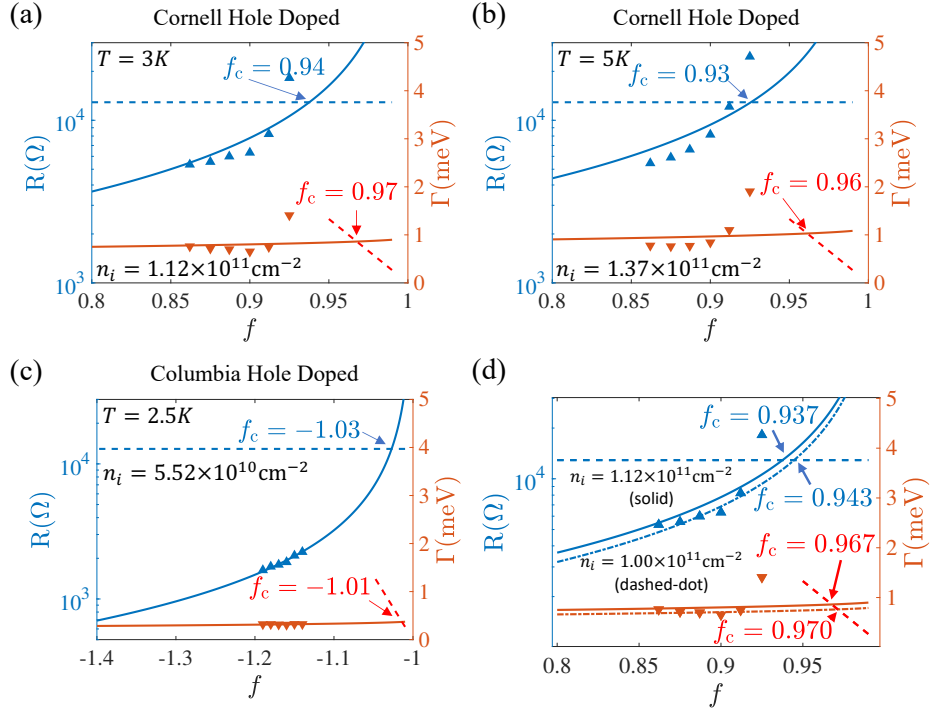

FIG. S1. The best fit to the finite temperature resistivity of the hole-doped TMD samples from the Cornell group at (a) T=3K and (b) 5K and (c) the Columbia group at 2.5K. The notations are the same as in Fig. 1-3 of the main manuscript. Here $n_i$ denotes the best fit impurity density. (d) Same as (a) but here we add the results (dashed-dot lines) calculated with $n_i = 1.00 \times 10^{11} \text{cm}^{-2}$, which is smaller than the best fit impurity density of (a) approximately by 10%. Note that the filling factors obtained for both impurity densities are almost identical.

In Fig. S1 we provide our best fits to the experimental resistivity measured at low finite temperatures where the resistivity measurement is less prone to error. While the overall fit quality is highly improved compared to the ones present in the main manuscript [c.f., Fig. 1(c) and Fig. 2(c)], the fit to the data point at $f = 0.925$ close to the critical filling factor ($f_c = 0.93 \sim 0.94$) in Figs. S1(a) and S1(b) is noticeably worse than the others because the Boltzmann theory does not apply very near the transition where localization and percolation effects are not negligible. Our fits here are obtained using the zero-temperature Boltzmann theory. Even though we may get more accurate results if we include finite temperature effects into our theory [23] (throughout the Supplemental Information reference numbers refer to the reference list in the main text), given that $T/T_F$ is small ranging from $0.01 \sim 0.1$ at given temperatures in Figs. S1(a)-(c), finite temperature effects are not significant enough to alter our conclusions, as shown in the following: At low temperature, the resistivity can be approximately written as [23]

$$\rho(T) \approx \rho_0 \left(1 + \frac{2x}{1+x}\frac{T}{T_F}\right) \tag{S1}$$

where $x = q_{\text{TF}}/2k_F$ with $q_{\text{TF}}$ being the Thomas-Fermi wavevector, and $\rho_0$ is the zero-temperature resistivity. Using this equation and the parameters corresponding to Figs. S1 (a)-(c), one can see that finite temperature effects increase

the resistivity by up to 10% or a bit more (i.e., $\frac{\rho(T)-\rho_0}{\rho_0} = \frac{2x}{1+x}\frac{T}{T_\mathrm{F}} \approx 0.01 \sim 0.1$). Since the resistivity is linearly proportional to the impurity density [see Eq. (1) of the main text], this calculation implies that the best-fit impurity densities obtained in Fig. S1 are overestimated roughly by 10% or a bit more than the impurity density that one would get with the finite temperature Boltzmann transport theory. However, this small overestimation of the impurity density barely affects our results qualitatively and thus do not alter our conclusions, as shown in Fig. S1(d). The fact that the finite temperature fit, avoiding the error-prone extrapolation to $T=0$, gives even better quantitative agreement to the experimental results shows that our theory captures the essential physics of the experiment, reinforcing the role of disorder in the experiments.

## FIELD INDUCED MIT FOR DEVICE 2

We provide our theoretical results in Fig. S2 for the field induced MIT as reported for Device 2 in Ref. [8] of the Cornell data. The six panels in Fig. S2 correspond precisely to the 6 panels in Fig. 3 of the main text except Fig. S2 is for Device 2 of Ref. [8] and Fig. 3 of the main text is for Device 1 of Ref. [8].

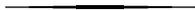

   

\end{document}